\documentclass[11pt,a4wide]{article}

\usepackage{amssymb}
\usepackage{amsmath}
\DeclareMathOperator{\acos}{acos}
\DeclareMathOperator{\atanh}{atanh}
\DeclareMathOperator{\cbrt}{cbrt}

\usepackage[hidelinks]{hyperref}
\newcommand*{\email}[1]{\href{mailto:#1}{\nolinkurl{#1}} } 
\usepackage{url}
\usepackage{graphicx}

\title{A framework to test interval arithmetic libraries and their IEEE 1788-2015 compliance}

\author{Luis Benet\(^1\), Luca Ferranti\(^2\), Nathalie Revol\(^3\) \\
{\small \([1]\): Instituto de Ciencias F\'isicas, Universidad Nacional Aut\'onoma de M\'exico, Cuernavaca, Mexico}\\
{\small \([2]\): University of Vaasa, Vaasa, Finland}\\
{\small \([3]\): INRIA - LIP (UMR 5668, ENS Lyon, University Lyon 1, Inria, CNRS), Lyon, France}
}

\date{Corresponding author: Nathalie Revol, INRIA - LIP, ENS de Lyon, 69364 Lyon Cedex 07, France. \email{Nathalie.Revol@inria.fr}}

\begin{document}
\maketitle
\begin{abstract}
As developers of libraries implementing interval arithmetic, we faced the same difficulties when it comes to testing our libraries.
What must be tested? How can we devise relevant test cases for unit testing? How can we ensure a high (and possibly 100\%) test coverage?
Before considering these questions, we briefly recall the main features of interval arithmetic and of the IEEE 1788-2015 standard for interval arithmetic.
After listing the different aspects that, in our opinion, must be tested,
we contribute a first step towards offering a test suite for an interval arithmetic library.
First we define a format that enables the exchange of test cases, so that they can be read and tried easily.
Then we offer a first set of test cases, for a selected set of mathematical functions.
Next, we examine how the Julia interval arithmetic library, IntervalArithmetic.jl, actually performs to these tests.
As this is an ongoing work, we list extra tests that we deem important to perform.

\noindent {\bf Keywords:} Unit tests for interval arithmetic libraries ; Test cases for interval arithmetic ; Testing IEEE 1788-2015 compliance \\
\end{abstract}

\section{Introduction}
\label{sec:intro}

\subsection{Context}
\label{subsec:context}
Interval arithmetic is an arithmetic that operates on intervals, that is, on sets of the form \([\underline{x},\bar{x}]\) rather than on numbers. The set \([\underline{x},\bar{x}]\) is an interval if \( \underline{x} \in \mathbb{R}\), \(\bar{x} \in \mathbb{R}\) and \(\underline{x} \leq \bar{x}\).
The result of the arithmetic operation \(\Diamond\) on the intervals \([\underline{x},\bar{x}]\) and \([\underline{y},\bar{y}]\) (if \(\Diamond\) takes two arguments), denoted by \([\underline{x},\bar{x}] \Diamond [\underline{y},\bar{y}]\), is the smallest (for inclusion) interval that contains the set \( \{ x \Diamond y, \: x \in [\underline{x},\bar{x}], \: y \in  [\underline{y},\bar{y}]\} \).
For usual operations, there are explicit formulas that express this definition in a more amenable way, for instance:
\[
[\underline{x},\bar{x}] + [\underline{y},\bar{y}] = [\underline{x} + \underline{y},\bar{x} + \bar{y}] ,
\]
\[
[\underline{x},\bar{x}] - [\underline{y},\bar{y}] = [\underline{x} - \bar{y},\bar{x} - \underline{y}],
\]
and
\[
[\underline{x},\bar{x}] \times [\underline{y},\bar{y}] = [ \min (\underline{x}\times\underline{y}, \underline{x}\times\bar{y},\bar{x}\times\underline{y},\bar{x}\times\bar{y}), \max(\underline{x}\times\underline{y}, \underline{x}\times\bar{y},\bar{x}\times\underline{y},\bar{x}\times\bar{y})].
\]
The fundamental principle of interval arithmetic is that, even if the input arguments are known with uncertainty, such as \(\pi \in [3.1415, 3.1416]\), or if the input arguments are only known to belong to a given set,
the computed result encloses the exact result (such as \(\sqrt{2 \pi} \in [ 2.506628274631000, 2.506628274631001]\)) or it encloses the range of the operation over the whole input sets (such as \( [0.5, 1.5]\times \sin (\sqrt{[2,4]}) \subseteq [0.4546, 1.5] \)).
This fundamental principle is sometimes expressed as "Thou shalt not lie: the sought result cannot be outside the computed result." Indeed, interval arithmetic offers a guarantee on the computed results, as they enclose the sought results.

However, usually interval arithmetic cannot be employed naively by replacing, in a given mathematical expression or program, all numbers by intervals and all operations by their interval counterparts.
As a matter of fact, interval computations are prone to overestimation, which can become so prominent as to hinder the relevance of the result.
A (stupid but archetypal) example is the evaluation of the expression \(x - x + x - x + x - x + x\), which is mathematically equivalent to \(x\) but, when evaluated using interval arithmetic with an interval of width \(w\), evaluates to an interval (containing the input interval) of width \(7w\).
Interval {\em analysis} is thus the (fine) art of devising expressions and programs that benefit from the guarantees provided by interval arithmetic and that determine tight enclosures of the results.
For a given function \(f\) and an interval \([\underline{x},\bar{x}]\), its goal is to determine an enclosure of the range \(f([\underline{x},\bar{x}]) = \{ f(x) \: : \: x \in [\underline{x},\bar{x}]\}\) without a too large overestimation.

In this work, we do not address interval analysis but we focus on libraries that implement interval arithmetic.
First, the definition of an admissible interval, for a library, must be clearly stated.
Depending on the underlying mathematical theory, unbounded intervals, the empty interval or intervals \([\bar{x},\underline{x}]\) with \(\bar{x} \geq \underline{x}\) are valid intervals or not.
In an attempt to clarify and to unify the definitions among the interval community, a standard for interval arithmetic has been elaborated, namely the IEEE 1788-2015 standard~\cite{IEEE-1788-2015}.
This standard first defines so-called "common intervals" which are intervals of the form \([\underline{x},\bar{x}]\) with \(\underline{x}\in\mathbb{R}\), \(\bar{x} \in \mathbb{R}\) and \(\underline{x} \leq \bar{x}\), that is, connected, closed and bounded sets of \(\mathbb{R}\).
The standard provides hooks to incorporate different mathematical theories, called "flavors".
For the time being, the only flavor defined by the standard is the set-based flavor, based on the set theory; in this flavor, \(\emptyset\) and unbounded intervals are valid intervals.
Furthermore, for the set-based flavor, the standard defines a set of decorations that provide additional information on the computation that has led to the interval under consideration: there is a decoration attached to each interval.
Let us consider the example of the interval \([0,1]\) that has been obtained as the result of \(\sqrt{[-2,1]}\): \(\sqrt{[-2,1]}\) is defined in the set-based flavor as \(\sqrt{ ([-2,1] \cap \mathrm{Dom}_{\sqrt{~}})}= \sqrt{[0,1]}=[0,1]\), where \(\mathrm{Dom}_{\sqrt{~}}\) denotes the domain of \(\sqrt{~}\), namely \([0,+\infty)\).
It is useful, in particular to avoid an incorrect application of Brouwer theorem (see explanations and more details in the standard~\cite[page 16]{IEEE-1788-2015}), to check whether such a intersection with the domain of the operation (of \([-2,1]\) with \([0,+\infty)\), resulting in \([0,1]\) in our example) has occurred.
The role of the decoration, attached to the result \([0,1]\), is to convey such information.
Libraries implementing interval arithmetic must clearly define upon which mathematical theory they are based; for instance they can claim compliance with the IEEE 1788-2015 standard.

These libraries are based on an arithmetic on numbers, to be able to compute for instance 
\(\underline{x} + \underline{y}\) and \(\bar{x} + \bar{y}\) when they perform the interval addition
\([\underline{x},\bar{x}] + [\underline{y},\bar{y}]\).
Often, the underlying arithmetic is the floating-point arithmetic as defined by the IEEE 754-2008 standard~\cite{IEEE-754-2008}, but not necessarily: arbitrary-precision floating-point arithmetic or exact rational arithmetic can also be employed.
Our focus is on libraries implementing interval arithmetic, and more specifically on testing these libraries.

Building tests along the development of a library is an important step towards gaining confidence in the quality of this library.
In this work, we focus on unit tests for libraries implementing interval arithmetic.
By a {\em unit test}, it is meant that each operation, or function, of the library, is tested individually; by contrast, benchmarks can be used to test a whole program that utilizes the library, and will not be addressed here. We refer to \cite{PPAM2022-TFSZKP} for an example of such a benchmark.
In our preliminary work~\cite{PPAM2022-RBFZ}, we presented our vision for a set of such unit tests, that can be used as a basis to check interval arithmetic libraries, or that can be incorporated as internal check procedures.

Let us first recall briefly the different categories we identified in~\cite{PPAM2022-RBFZ}, then we will sketch the content of this article, with an emphasis on our contributions.

\subsection{Which tests?}
\label{subsec:which-tests}
\subsubsection{Definition of unit tests}
We recall here our definition and notations for unit tests, following~\cite{PPAM2022-RBFZ}.
Let us denote by \(\cal{L}\) the tested library, and interval quantities using boldface: \(\mathbf{x}\), \(\mathbf{y}\). Here \(\mathbf{x}\) can be an interval or an interval vector, that is, a vector whose components are intervals, or any other interval quantity such as a matrix, even if this latter case does not occur yet in the presented work.
We focus here on unit tests, for a function, denoted by \(f\).
A unit test case is a pair composed of the input argument \(\mathbf{x}\) and the expected output \(\mathbf{y}\).

First, the output \(\mathbf{y}\) must be the tightest representable interval enclosing \(f (\mathbf{x}) \); otherwise a very accurate library could compute \(\mathbf{z}\) such that \( \mathbf{z} \subsetneq \mathbf{y} \) and still \( \mathbf{z} \supseteq f(\mathbf{x}) \).
To ensure the tightness property for \(\mathbf{y}\),
typically, one computes the endpoints \(\underline{y}\) and \(\bar{y}\) of \(\mathbf{y}=[\underline{y}, \bar{y}]\) using a precision (that is, a number of binary -- or decimal in case the library uses decimal arithmetic -- digits used for the computations) higher than the computing precision of \(\cal{L}\).
Let us illustrate this with \(\cal{L}\) using {\tt binary64} floating-point numbers, on the (simple) example of the exponential function. We assume that the given floating-point implementation of \(\exp\) does not provide correct rounding, but that, for any precision \(q\), it satisfies \(RD_q(\exp(x)) \leq \exp(x) \leq RU_q (\exp (x)) \), for \(RD_q\) rounding downwards, and \(RU_q\) rounding upwards, both in precision \(q\).
Given \(\mathbf{x} = [ \underline{x}, \bar{x}]\), to compute the infimum \(\underline{y}\) of \(\mathbf{y}\), one gets the approximations \(RD_q(\exp(\underline{x}))\) and \( RU_q(\exp(\underline{x}))\) of \(\exp(\underline{x})\) in high precision \(q\), and finally round them downwards in the target precision, \(53\) for {\tt binary64}.
If \(RD_{53}(RD_q(\exp(\underline{x})))\) and \(RD_{53}(RU_q(\exp(\underline{x})))\) are equal, then they are the sought value \(\underline{y}\) for the infimum of \(\exp(\mathbf{x})\).

Once the test cases are devised, what conclusions can be drawn from the comparison between \(\mathbf{y}\) and \(\mathbf{z}\) the result computed by \(\cal{L}\)? Requiring equality may be too demanding. Inclusion is required, but we want to dismiss, for instance, an implementation of \(\sin\) that returns \([-1,1]\) for any argument. If the accuracy of the function is given in the specification of the library, how can it be used and checked?

A last general recommendation is to incorporate the following procedure: if \( (\mathbf{x}, \mathbf{y})\) is given as a pair of input and output, pick at random (a reasonably large number of) values \(x \in \mathbf{x}\) and check whether \(f(x) \in \mathbf{y}\) with \(f(x)\) computed by the underlying arithmetic, and whether \(f([x,x]) \subseteq \mathbf{y}\). Those tests are disconnected from the knowledge of the implementation of \(f\) and may hit a zone not considered (forgotten) in the development of \(f\), such as an overlooked quadrant for a trigonometric function.
\\

We now detail the different categories of unit tests we recommend. First, there are unit tests that can be used to check any interval arithmetic library. Then some tests are designed to check specifically IEEE 1788-2015 compliance. Finally, each library has its own unique features that must also be tested. Let us detail these three categories.

\subsubsection{Tests common to all libraries}
As every interval arithmetic library implements usual arithmetic, mathematic and interval-specific functions, unit tests must check these functions.
The set of unit tests must encompass:
\begin{itemize}
\item {\bf easy test cases:} these are useful especially during the early development phase, to detect and correct ``obvious'' bugs;
\item {\bf special and exceptional values:} the handling of such values is usually not difficult, but it is easy to omit one of them;
test cases must then be exhaustive, to test every possible configuration;
\item {\bf cornercases:} for instance when a function is not monotonic, such as the \(\sin\) function, test cases must give examples
where the interval includes, or not, the optimum; cases where one of the endpoints is close to this optimum are difficult ones
and must appear in the set of test cases. Another class of cornercases is related to the specificities of the underlying floating-point
arithmetic, if it is used to implement the interval arithmetic: it includes subnormal values and hard-to-round cases.
The test cases in this category heavily depend on the underlying arithmetic, such as floating-point arithmetic for rounding issues,
and its precision (e.g. {\tt binary32} or {\tt binary64}) for subnormal values.
\item {\bf interval-specific functions:} test cases must also target specific functions, such as the "union" of two intervals,
which returns the convex hull of the union, or the radius of the interval~\cite{Goualard2014}.
\item {\bf input and output:} I/O functions must be thoroughly tested, in particular because the most unexpected inputs can be given.
\end{itemize}

\subsubsection{Testing IEEE 1788-2015 compliance}
We do not detail here the content of the IEEE 1788-2015 standard~\cite{IEEE-1788-2015} for interval arithmetic. We only allude to the features relevant
to the development of test cases.
\begin{itemize}
\item As I/O are very precisely defined by the standard, the different possible inputs and outputs must be tested according to
these specifications.
\item IEEE 1788-2015 is designed to allow for several mathematical theories to be the theoretical foundations for the implemented
interval arithmetic: each theory is called a {\em flavor}. As the only standardized flavor so far is the set-based flavor,
test cases for the corresponding flavor must be given: they encompass unbounded intervals as well as empty intervals.
\item The standard gives two lists of mathematical functions, the required ones (such as addition or $\sin$) and the recommended
ones (such as $\textrm{log2p1}$).
The tests should include test cases for all of these functions, but should not fail in the event a
recommended function is not implemented.
\item Decorations are a rather unique feature of the IEEE 1788-2015 standard: they provide some additional information on the
history of the computation that yields the current result. Test cases must include all possible decorations for the inputs,
in order to check that the decoration of the output is correctly determined.
\item The IEEE 1788-2015 standard specifies different accuracies, from {\em tight},
where each result is the tightest representable
interval that encloses the exact result, to {\em valid}, where the computed result must simply include the exact one.
For the set-based flavor, an additional mode is the {\em accurate} mode, which is very precisely defined by the standard, but corresponding values require care to be determined.
Test cases must include all of these possibilities, even if the tested library provides only some of them; its behavior
according to the claimed accuracy must be tested.
\end{itemize}

\subsubsection{Tests specific to some libraries}
Interval arithmetic libraries may use different representations for the interval, e.g. by their endpoints as in MPFI~\cite{MPFI}
or by their midpoint and radius as in Arb~\cite{ARB}.
They may rely on usual floating-point arithmetic as Octave interval package~\cite{OctaveIntervalPub}, or arbitrary precision
floating-point arithmetic as MPFI, or exact rational arithmetic as JInterval~\cite{JIntervalPub}.
Some libraries use usual floating-point arithmetic but avoid changing the rounding mode, as Filib~\cite{FilibPub} or JuliaInterval~\cite{ValidatedNumericsPub},
or as GAOL~\cite{Goualard2005} which can use either rounding-to-nearest exclusively, or rounding-upwards exclusively.
Each specificity must be tested through several different test cases.
Typically, several examples using largely varied precisions must be tested for a library using arbitrary precision.
Examples which are cornercases for EFT (Error Free Transform)~\cite[Chapter 4]{HandbookFP} must be included to test libraries that are based
on rounding-to-nearest and that use EFT to round outwardly their results.

However, since these tests are very specific to a given library, either their presence in a database is optional,
or running them should not be systematic.

\subsection{Our contributions}

Our contributions can be summarized as follows.

\begin{itemize}
    \item We introduce a JSON schema to describe interval tests in Section~\ref{sec:JSON}. This allows the tests to be easily integrated in existing libraries.
    \item We present an extensive database of tests in Section~\ref{sec:new-version}. This builds on previous efforts such as ITF1788 presented in Section~\ref{subsec:platforms}, though we expand it with new test cases.
    \item In particular, we generate tests whose intervals have hard-to-round floating point numbers. These cases were generated using the program {\tt BaCSeL} and are particularly important to ensure that an interval implementation is robust to floating-point rounding issues. How these tests have been generated is detailed in Section~\ref{subsec:gener-test-cases}.
    \item Finally, we run our tests on the \textsc{IntervalArithmetic.jl}. As developed in Section~\ref{sec:results}, we show how our newly generated hard-to-round cases can help unveil hard to detect bugs, making hence our contribution valuable for interval arithmetic library developers.
\end{itemize}

\subsection{Related work}

Our work relates to unit testing, that has largely been studied and discussed. We sum up the main points below, before introducing existing platforms offering unit testing of interval arithmetic libraries.

\subsubsection{Unit testing}
In the field of software testing, various approaches are possible and are usually applied one after the other.
Unit testing is usually the first step, it consists in testing each developed function independently of the other ones.

Of course, unit testing is useful to detect bugs. However, a function that successfully passes the provided unit tests is not guaranteed to be correct.
An approach to guarantee correctness would be to use formal proof, but this is not in the scope of this work. We refer to Gappa~\cite{Gappa} and \cite[Section 13.3]{HandbookFP}, and Flocq, based on Coq.Interval~\cite{Flocq}.

Unit testing is usually the first step to test a library.
It is then followed by integration testing that tests whether a whole program behaves properly, and thus whether the various functions of the program perform properly together.
Relevant tests for this phase of integration testing are benchmarks such as the one given in~\cite{PPAM2022-TFSZKP}.

The field of software testing has developed various metrics to quantify the quality of the tests, such as code coverage or branch coverage~\cite{ZhuHallMay1997}. These metrics are not relevant for our work. Indeed, our tests are oblivious to the implementation, they only focus on the semantics of the computations: the image of an interval by the sine function has a mathematical meaning, which is decorrelated from the actual implementation of the tested sine function.

Finally, we would like to emphasize the conclusions of Ellims, Bridges and Ince~\cite{EllimsBridgesInce2004} about unit testing:
unit tests allow developers to detect, at an early stage of the development, many errors, for a cost per error, measured in hours of human work, which is about a tenth of the cost of detecting an error during the integration phase.
\\

\subsubsection{Existing platforms}
\label{subsec:platforms}

Although interval arithmetic was standardized in 2015, no official test suite to test for compliance was provided. Traditionally, each library relied on its own in-house tests.
However, Kuliamin~\cite{Kuliamin2007} lists all kind of input values needed to test thoroughly a floating-point arithmetic library. For the exponential function for instance, this amounts to 15,000 test cases. 
For an interval-valued function, one gets \(15,000 \times 15,000\) possible inputs. This may be extreme, but it illustrates the need to share unit tests, rather than letting each developer of an interval arithmetic library devise her/his own tests.
\\

The first effort in this direction was the JInterval P1788 Test Launcher~\cite{JIntervalLauncher} based on the JInterval library~\cite{JIntervalPub} for interval arithmetic in Java.
For each tested library, a wrapper must be written to enable to call the operations and functions of this library.
Wrappers are available for \texttt{Profil/BIAS}, \texttt{boost/interval}, \texttt{C-XSC}, \texttt{Filib}, \texttt{libieeep1788}, \texttt{libMoore}, \texttt{MPFI}.
The launcher reads tests from plain text files of simple human-readable format and writes the results computed using tested library and JInterval and their relation into a plain text report. Test set included in the JInterval P1788 Test Launcher consists of over 14,000 tests which partly originated from \texttt{libieeep1788}, \texttt{Filib}, \texttt{libMoore} while the other ones are original.
\\

Another  attempt to introduce a systematic testing framework for interval arithmetic was the Interval Testing Framework 1788 (ITF1788) \cite{ITF1788Pub}.
Developers of ITF1788 introduced a domain specific language called ITL (Interval Testing Language) that could be used to express tests for interval arithmetic.
ITF1788 is a Python engine that inputs ITL-files and converts tests into code for the specified language, test framework and library.
However, while being an elegant language, this ITL solution presents a few drawbacks.
First, being an ad-hoc language, it also requires an ad-hoc parser for parsing the tests into a structured representation and/or produce code in the target language.
To overcome this, we choose to use the JSON format, which is widely used in software engineering and comes with available parsers.
Secondly, the proposed language lacked flexibility to test different precision or to test for both accurate and tight mode; our proposal offers the required flexibility.

After the introduction of ITL, a first attempt to produce an exhaustive test-suite was presented in \cite{ITF1788Codes}.
This work both collected tests from existing interval libraries, namely \texttt{C-XSC}, \texttt{FILIB}, \texttt{MPFI}, \texttt{libieeep1788}, and produced several new test cases for features introduced by the standard, such as decorated intervals.
In our work, we build on top of this previous attempt, by porting the existing ITF1788 tests into JSON format and enriching the test suite with new test cases for the accurate mode, as well as new tests for different precision and tests for hard-to-round cases.

\subsection{Parallel or distributed computations}
Testing correctness of interval libraries under parallel behavior is important, especially for libraries that handle floating-point error by changing the rounding mode.
In several common processors, the rounding mode of the floating processing unit is controlled by a global state in the unit registers. If not handled with care, manually changing the rounding mode in parallel codes can lead to race conditions and undefined behavior.
To test library soundness under multithreaded execution, the tests presented here could be run in parallel. In addition to speeding up the tests execution, this could also be used to test that correct rounding is preserved under parallel execution.

Parallelism was also used to generate the hard-to-round cases: the code that determines hard-to-round cases, called {\tt BaCSeL}, uses multithreading to explore in parallel many subintervals. Some timings are given in~\cite{Core-math2022}, that justify the need for large computing power and parallelism.

\section{A specification language for interval arithmetic testing}
\label{sec:JSON}

We propose to specify the tests for our framework using the JSON format~\cite{JSON}.
To quote the introduction to JSON, ``JSON (JavaScript Object Notation) is a lightweight data-interchange format.
It is easy for humans to read and write. It is easy for machines to parse and generate.
It is based on a subset of the JavaScript Programming Language\([\ldots]\)
JSON is a text format that is completely language independent but uses conventions that are familiar to programmers of the C-family of languages\([\ldots]\)''
This being a popular format, it can be easily read seamlessly by several languages, while still being human-readable. JSON being the most common format for web-development, this could potentially allow to develop a web-interface to test interval libraries.

Each test-case is a JSON object with the following fields:

\begin{itemize}
\item function: specifies the name of the function-under-test (FUT), must be as named in the standard;
\item precision: the number of bits of precision used;
\item input: specifies the input, see below for description of how the input is specified;
\item output: an object with two fields: \textsc{tight} (required), and \textsc{accurate} (optional), specifying the result for the tight and accurate modes.
For functions returning a number or a boolean (e.g. a function returning the midpoint of the interval or a comparison), only the tight
mode shall be provided. The format corresponding to each mode is similar to the format used for the input.
\end{itemize}

Both inputs and outputs must have a field  \textsc{type}, which can have one of the following values: number, interval, boolean or string.
For types \textsc{boolean}, \textsc{number} and \textsc{string}, the only required additional field is \textsc{val}, which must contain the value.
If the type is \textsc{interval}, then the object has two compulsory fields \textsc{inf} and \textsc{sup}, specifying the lower and upper endpoints of the interval,
respectively. Intervals can have an optional third field, \textsc{dec}, specifying the decoration.

The following code snippet shows an example of a function.

\begin{verbatim}
  {
    "function": "atanh",
    "precision": 23,
    "input": [
      {
        "type": "interval",
        "inf": "-0xf.fe1e00@-1",
        "sup": "0xf.fe1e00@-1"
        "dec": "com"
      }
    ],
    "output": {
      "tight": {
        "type": "interval",
        "inf": "-0x4.305fa0@0",
        "sup": "0x4.305fa0@0",
        "dec": "com"
      },
      "accurate": {
        "type": "interval",
        "inf": "-0x4.306830@0",
        "sup": "0x4.306830@0",
        "dec": "com"
      }
    }
  }
\end{verbatim}

In the above code snippet, the \textsc{inf} and \textsc{sup} entries are hexadecimal numbers whose exponent is in base 16.\\

\section{New test cases}
\label{sec:new-version}

One of our contributions is to devise and to share publicly a suite of unit test cases, for arithmetic and mathematical operations and functions. Ideally, the whole set of functions recommended by the IEEE 1788-2015 standard would be covered.
In this work, we share the set of test cases which have already been gathered within ITF-1788~\cite{ITF1788Codes} (which was a collection of test cases from various existing libraries: C-XSC, FILIB, MPFI, libieeep1788 in particular) and new ones.

Our goal is to offer and to share test cases in all categories in Section~\ref{subsec:which-tests}. In what follows, we do not develop the first categories, and we focus on hard-to-round cases because they took most of our time.

So, here are just a few indications about the generation of the other test cases.
\begin{itemize}
\item {\bf Easy cases} are small, usual values. The choices may be arbitrary, but they offer the advantage to be easy to read and there are quantities that are easy to check, such as sign or magnitude; examples include \(\sin 0\) or \(\exp 0\) or \(\cbrt(-64)\), where \(\cbrt\) stands for cubic root.
\item {\bf Special cases} are also included, they cover both exact cases and floating-point special values.
For transcendental functions, there are very few "exact" cases, apart from \(0\) (as argument or as result): this is true for functions deriving from the exponential as stated by Lindemann-Weierstrass theorem~\cite{Lindemann,Weierstrass}, and for functions deriving from the logarithm as proven by Baker's theorem~\cite{Baker} (see also Wikipedia for a more accessible presentation of these theorems).
Special values, for floating-point arithmetic, are \(0\) (with its two signed representations \(-0\) and \(+0\)) and infinities, both as inputs and as outputs, as in \(\exp(-\infty)\) or \(\atanh(1)\). Each of these special values can appear as the left endpoint or right endpoint of an interval, or both.
\item {\bf Exceptions.}
We must then test for instance \( \exp[-\infty, -\infty]\): in this case, the input argument is ill-formed and can be considered either as the empty set (if the flavor so permits) or as an error.
This is an example of an exceptional behavior. Other examples of exceptional values that must be tested include {\tt NaN} as one or both endpoints. We did our best to include all such possibilities in our test cases.
\item {\bf Cornercases.} The bulk of our contribution in this section consists in the determination of cornercases, in particular in the determination of values which are hard to round. Indeed, these cases are difficult to evaluate accurately and thus a library is most likely to fail. In what follows, we develop our approach to determine these cornercases.
\end{itemize}

\subsection{Generation of the test cases}
\label{subsec:gener-test-cases}
\subsubsection{Generation of hard-to-round cases}

For a floating-point number \(x\), 
its image by a function \(f\): \(f(x)\) is written in the working base \(B\) as \(f(x) = \pm y_0. y_1\ldots y_{p-1} y_p \ldots . B^e\) for some exponent \(e\), with an infinite sequence \( (y_i)\) of bits (if \(B=2\)) or digits (if \(B=10\)).
Without loss of generality, let's assume that \(B=2\).
The floating-point number \(\hat{y}\) that represents \(f(x)\) in the floating-point format has \(p\) bits, it is the rounding of the value \(f(x)\).

If, for instance, the rounding mode is towards \(+\infty\) and the approximation of 
\(f(x)\) using \(n \geq p\) bits is \(\pm y_0. y_1\ldots y_{p-1} 0\ldots 0 .B^e\), i.e. \(y_p = \ldots = y_{n-1} =0\) and the error of the approximation is bounded by \(B^{e-n+1}\) which is the weight of the last bit of this approximation,
then one doesn't know how to round \(f(x)\), indeed there are two possible candidates: \(\pm y_0. y_1\ldots y_{p-1} . B^e\) and \(\pm (y_0. y_1\ldots y_{p-1} +B^{-p+1}). B^e\).
When the difference between \(p\) and \(n\) is large, one must evaluate \(f(x)\) with a large precision \(\geq n\) to be able to round \(f(x)\) correctly into \(\hat{y}\): \(x\) is said to be {\em a hard-to-round case} for \(f\) and the floating-point format. In what follows, we do not focus on the hardest-to-round cases, that is, the values of \(x\) that require the largest value of \(n\), but only on sufficiently hard-to-round ones.
This problem is well-known to people who produced (by hand, before computers existed) tables for mathematical functions, hence its name: the Table-Maker Dilemma, TMD in short.

The hard-to-round cases we propose have been determined using {\tt BaCSeL}~\cite{BaCSeL}, a program of about 5,000 lines of code in C, written and maintained mostly by P. Zimmermann. 
It is primarily intended for a personal use, with a concise documentation given as a README and an installation that requires GMP, MPFR, FPLLL and OpenMP.
It has been developed since 2020 and it aims at determining hard-to-round cases for several mathematical functions, over specific subdomains.
It is based on several algorithms that rely on LLL to prune easy cases and to focus on hard ones.
The computing basis is either \(2\) or \(10\), in what follows we focus on the basis \(2\).
The list of available functions contains \(\exp\), \(\exp_2\), \(\log\) and \(\log_2\), \(\sin\), \(\cos\), \(1/x\), \(\sqrt{x}\), \(1/\sqrt{x}\), \(\sqrt[3]{x}\), \(\acos\) where the subscript indicates the basis of the exponential or the logarithm: \( \exp_2(x) = 2^x\) and \(\log_2 x = \log x / \log 2\).
Each subdomain is split into smaller intervals, that can be further subdivided to reach the required accuracy.
These smaller intervals are explored in parallel, using OpenMP to distribute the work.
Each function is internally approximated by a Taylor expansion on each small interval.

More precisely, the {\tt BaCSeL} determines all hard-to-round cases for a given function \(f\) over an interval \([t_0,t_1]\).
The interval is such that \(t_0 < t_1\) and \(t_0\), \(t_1\) belong to the same binade, in other words there exists an integer \(n\) such that \(2^{n-1} \leq t_0 < t_1 \leq 2^n\) (and the interval \([2^{n-1}, 2^n)\) is called a {\em binade}). It is required that the image of \([t_0,t_1]\) by \(f\) does not overlap several binades.
The "hardness-to-round" is given by the user, it corresponds to the number \(n-p\) in our definition of a hard-to-round number.
The precision of the input arguments and of the evaluated results must be specified by the user, as well as the computing precision needed to evaluate accurately enough the function, in order to determine whether an argument \(x\) is a hard-to-round case; in particular this computing precision must be larger than \(n\).
Several other values must be fed to {\tt BaCSeL} so that it determines hard-to-round cases. The short description given here gives a hint of the number of trial-and-errors required to choose proper intervals and precision, in order to get a positive and reasonable number of hard-to-round cases.
In other words, some expertise along with computational power are useful to determine hard-to-round cases.

\subsubsection{Computation of the expected results}
Once the input values are listed, either easy ones, or special ones, or cornercases, then the expected result must be computed.
In this work, we rely on MPFI~\cite{MPFI} to evaluate the given function \(f\) over the input interval \(\mathbf{x}\).
MPFI is designed to return \(f_{\mathrm{tightest}} (\mathbf{x})\), the tightest interval enclosing the mathematical image \(f(\mathbf{x})\), with endpoints that are floating-point numbers at the prescribed precision. It thus gives the target result for the "tight" mode of the IEEE 1788-2015 standard.

MPFI is also used to give the largest admissible result for the "accurate" mode of the set-based flavor.
This mode is defined as follows in the standard~\cite{IEEE-1788-2015}. 
Let \(x\) be a floating-point number and let us denote by \(\mathrm{nextUp}(x)\) the smallest floating-point number strictly larger than \(x\), it can be \(+\infty\). Similarly, let us denote by \(\mathrm{nextDown}(x)\) the largest floating-point number strictly smaller than \(x\), it can be \(-\infty\).
Finally, let us denote by {\tt nextOut} the function that returns, for an interval \(\mathbf{x}=[\underline{x}, \bar{x}]\) where \(\underline{x}\) and \(\bar{x}\) are floating-point numbers, the interval \([\mathrm{nextDown}(\underline{x}), \mathrm{nextUp}(\bar{x})]\).
The "accurate" mode requires that the evaluation of a function \(f\) over an interval \(\mathbf{x}\) is enclosed in
\(\mathrm{nextOut} ( f_{\mathrm{tightest}} (\mathrm{nextOut}(\mathbf{x})))\).
MPFI has been enriched with the \(\mathrm{nextOut}\) function, so as to determine this interval as well.

\subsubsection{Format for the input and the expected output}
In this work, we use the capability of both {\tt BaCSel} and MPFI to output their results in hexadecimal format: this enables writing and reading floating-point numbers without any conversion error from and to radix 10.

\subsection{Four mathematical functions as illustration: cubic root, exp, sin, atanh}

As the IEEE 1788-2015 standard for interval arithmetic requires over 50 mathematical operations and functions and recommends 16 additional ones for the set-based flavor, we will not detail the test cases for each function but we will focus on a few selected ones. The ones we choose here are the cubic root, the exponential, the sine and the hyperbolic arc-tangent.
These test cases can be found at \url{https://github.com/lucaferranti/IntervalArithmeticTests}.

\subsubsection{Test cases for {\tt cbrt}}
The cubic root is a relatively simple function, as it is algebraic, it is defined on the whole set of real numbers, and it exhibits a symmetry (indeed it is odd) which can easily be checked.
Furthermore, examples involving the cubic root are easy to check, at least approximately: first because it is monotonic, and also as multiplying each endpoint of the result by itself twice should land close to the original argument.
Last, because the cubic root is algebraic, many inputs yield an exactly representable result. Thus, exact cases are numerous.
Examples where the floating-point endpoints have exponents which are integer multiples of 3 are easily reduced to the former case.

\begin{figure}
\begin{center}
\includegraphics[width=65mm]{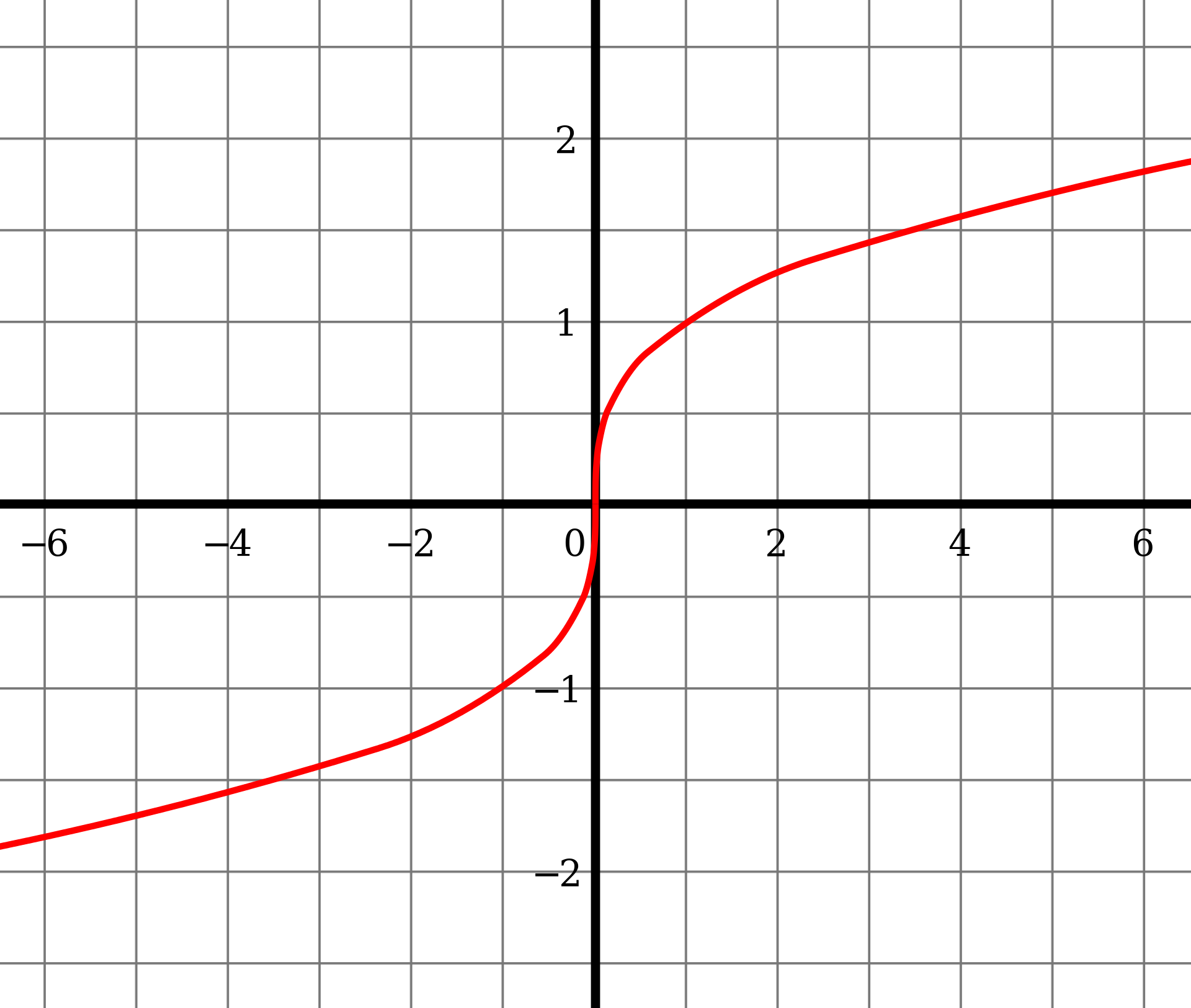}
\caption{Graph of the cubic root. (Credit: Krishnavedala - CC BY-SA 4.0.)} 
\label{fig:cbrt}
\end{center}
\end{figure}

In our test cases, we focus on the zone close to \(0\), where the function has a vertical tangent, see Figure~\ref{fig:cbrt}: since it is the place where the variations are the fastest, this is the zone where we expect numerous hard-to-round cases to lie.

\subsubsection{Test cases for exp}
Our other three functions are transcendental ones.
The simplest one is the exponential: it is a very well studied function, it usually comes with a high quality implementation.
It is also a "simple" function, in the sense that it is defined everywhere, it is positive, monotonic, it has only one "exact" value in the sense that only one floating-point value, namely \(0\), has an image which is exactly representable as a floating-point number, namely \(1\).
The domain on which it takes finitely representable values is bounded from above: for the {\tt Binary64} format, \(exp x\) overflows for \(x \geq 710\).

Our test cases include mostly hard-to-round values around \(0\), that have been determined using {\tt BaCSeL}.
This is a first step to assess the quality of the implementation of the exponential function.
Indeed, to evaluate \(\exp x\), the implementation usually involves a {\em range-reduction} step, where the argument \(x\)
is repeatedly divided by \(2\) until it becomes \(x' = x/2^n\) close to \(0\).
Then \(\exp x'\) is evaluated accurately.
Eventually \(\exp x\) is obtained as \(\exp x'\) squared \(n\) times: this last phase is called {\em reconstruction}.

Future work will consist in adding values:
\begin{itemize}
\item close, but smaller than the upper limit of the domain where \(\exp\) overflows;
\item large in absolute value, and negative, to check whether subnormal results are properly handled;
\item more varied in magnitude, to check the accuracy of the reconstruction phase as well.
\end{itemize}

\subsubsection{Test cases for sin}
Our second transcendental function is the sine function.
Again, there is only one "exact" value: \(\sin 0 = 0\).
This function is more difficult to implement, as it is non monotonic.
As it is periodic, the first step to evaluate \( \sin x\) is again a range-reduction phase: it consists in translating
\(x\) by the appropriate integer \(k\) multiple of the period \( 2 \pi\), to obtain \(x' = x -2k\pi\) close to \(0\).
Contrary to what happened for the exponential, there is no reconstruction phase for the sine function.

The range-reduction can be delicate: as \(\pi\) is transcendental, the implementation can require a very large computing precision to determine \(k\) and then to compute accurately \(x' = x -2k\pi\), especially when \(x\) is large.
We test such extremely large arguments \(x\) in our test cases.
The other test cases we propose are hard-to-round cases close to \(0\): these values are useful to determine whether
the evaluation of the sine is accurate, once the range-reduction phase is performed.

It is easy to check whether the results are correct, in the "tight" mode: for such tiny values, \(\sin x \sim x\) applies.
It is also easy to check whether the results are correct in the "accurate" mode: the input argument has been enlarged by one ulp at each endpoint, and the tight evaluation with this enlarged argument is again enlarged by one ulp at each endpoint,
the difference between the tight and the accurate evaluations are 2 ulps at each endpoint. As the values are given in hexadecimal format, this difference of 2 ulps at each endpoint is easy to check.

Future work will consist in adding more varied input arguments, both in sign and in magnitude, and also in adding intervals closer to an extremum \( \frac{\pi}{2} + k \pi\), either containing the extremum or not.
Some expertise in approximation theory is needed to identify floating-point arguments that are very close to an optimum; continued fractions are useful as illustrated in Chapters 4 and 10 of \cite{HandbookFP}.
Again, all our test cases have been determined using {\tt BaCSeL}.

\subsubsection{Test cases for atanh}
\begin{figure}
\begin{center}
\includegraphics[width=65mm]{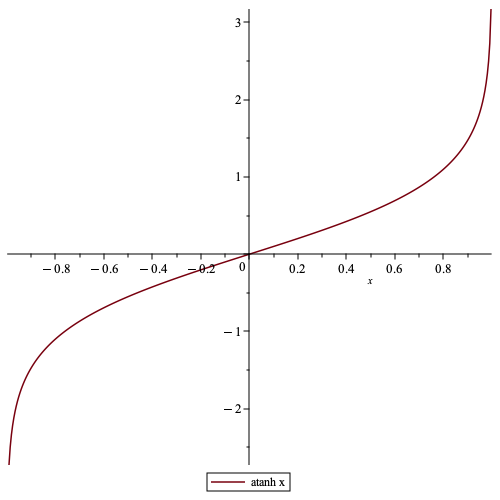}
\caption{Graph of the hyperbolic arc-tangent function.} 
\label{fig:atanh}
\end{center}
\end{figure}

Our last transcendental function is the hyperbolic arc-tangent \(\atanh\).
We choose this function because, in our experience, this function is usually implemented with much less accuracy than the two previous, more usual, ones.
Again there is only one "exact" value" \(\atanh 0 = 0\).
This function is rather simple to study: it is odd, monotonic, defined on a bounded domain \((-1,1)\), with infinite limits at the boundaries, see Figure~\ref{fig:atanh}:
\[ \lim_{x \rightarrow -1^+} \atanh x = -\infty \mbox{  and  } \lim_{x \rightarrow 1^-} \atanh x = + \infty .\]

Our test cases contain hard-to-round cases determined using {\tt BaCSeL}. 
We first had to integrate the hyperbolic arc-tangent function in the list of functions known by {\tt BaCSeL}.
Our goal was to cover several intervals bounded by successive powers of \(2\) and included in the domain of \(\atanh\),
including values close to the boundaries of the domain, however \(\atanh\) does not grow very rapidly and we did not find (yet) any value where the evaluation overflowed.

\section{Results with the Julia Interval library}
\label{sec:results}

The {\tt IntervalArithmetic.jl} package is written in the Julia programming language,
motivated by Julia's mathematical-style syntax, near C-performance, high-level
interactive usability, very robust type system with easy to exploit composability
properties.
These capabilities of the language permit to have parametric interval that depend
on the underlying floating-point representation, which can be extended to
incorporate the decoration system, the possibility
for different accuracy modes, and flavors. The package is on its way to be fully
conforming with the IEEE 1788-2015 Standard, with most tests of the ITL test suite
successfully passing.

IntervalArithmetic.jl implements interval-types in the form {\tt Interval$\{$T$\}$},
where {\tt T} refers to the underlying floating-point representation.
For the tight accuracy mode, the package
relies on the {\tt CRlibm} library~\cite{CRlibm} for correctly rounded {\tt Binary64},
and on the {\tt MPFR} library~\cite{MPFR} for {\tt Binary32} and
arbitrary precision. The {\tt CRlibm} library does not have all
functions that the IEEE 1788-2015 standard requires, for instance, $\atanh$ is not
included. In those cases, we use the MPFR library, and convert the final result
to the floating-point representation of the input.

The results discussed below use the {\tt 1.0-dev} branch, and are produced
with the {\it tight} mode of accuracy only. For this accuracy mode, we check the
identity of the result produced by the library and the hard-to-round results.
Below we do not report the tests that involve the empty set, since they are already
addressed by the ITL test suite.

All test cases for the $\exp$ and $\sin$ functions pass correctly. For these functions,
the hard-to-round tests we produced include examples for {\tt Binary32} and
{\tt Binary64} numbers only.
While these results are positive, it remains to produce and test hard-to-round
cases that involve other precisions than the usual formats.

For the $\atanh$ and $\cbrt$ functions, the results show tests that pass and others
that fail. In both cases, the tests that systematically pass correspond
to the {\tt Binary64} format. This statement is not trivial since none of these
functions is included in the CRlibm library.
The failing tests then involve the {\tt Binary32} format, or cases involving
different precision values for extended precision format.
The positive aspect of the failing tests that we have uncovered is that
some functions that have naively been extended, say to {\tt Binary32}, should be
handled more carefully.

\section{Conclusions and future work}
\label{sec:conclusions}

As interval arithmetic aims at producing verified computations that can be used for rigorous mathematical proofs, standardization and testing is a crucial topic.
Despite the standardization of interval arithmetic in 2015, little attention has been given to how testing should be implemented to ensure the library is sound and standard compliant.
In this paper, we reviewed the principal approaches and challenges of testing interval libraries. Building on previous work, we produced an open source collection of tests for interval arithmetic libraries.
This test suite is released in JSON format, which allows it to be easily integrated by different libraries in different languages.
In addition to collecting already existing tests, our test suite tackles challenges that were previously unaddressed.
First, our test suite contains test cases for the accurate mode described in the standard. This helps libraries developers to determine whether their results are tight, accurate or simply valid.
Second, using the BaCSeL program, we generated hard-to-round cases for some elementary functions. These are important cornercases to test the library robustness against floating-point rounding issues.

As future work, we plan to keep expanding our test suite with more hard-to-round cases both for {\tt Binary32} and {\tt Binary64}, for the whole set of functions that are either required or recommended by the IEEE 1788-2015 standard for interval arithmetic.
We also plan to expand our test suite to check more thoroughly specificities of the standard, such as the computation of decorations, or the handling of unbounded or empty intervals.

We also plan to test other libraries. For instance, we could test the accuracy of the ARB library, which uses a midpoint-radius representation for the intervals.
Indeed, while interval arithmetic is usually concerned with inf-sup format, the midpoint-radius format has proven itself an appealing alternative, as one can choose the appropriate trade-off between accuracy and efficiency.
Algorithms that use the midpoint-radius representation can be fast but at the expense of the accuracy, and this seems to be the case of ARB, at least for some operations, or they can be slower and use more complicated formulas to remain very accurate. It would be interesting to determine whether the results computed by ARB, once converted to a representation by endpoints, fall into the tight, accurate or simply valid category.

Finally, midpoint-radius arithmetic has received less attention and there has not been much effort in standardizing it. Future work will focus on generating test cases for midpoint-radius arithmetic, which can be another approach to assess the accuracy of a library as popular as ARB.

\subsection*{Acknowledgement}
Luis Benet acknowledges funding from IG-101122 DGAPA (UNAM) project. 

A CC-BY public copyright license has been applied by the authors to the present document and will be applied to all subsequent versions up to the Author Accepted Manuscript arising from this submission.

\end{document}